# Improved Power Counting and Fermi Surface Renormalization [*]

MANFRED SALMHOFER

*Mathematik, ETH Zürich, CH–8092 Zürich*

manfred@math.ethz.ch

**Abstract:**
The naive perturbation expansion for many-fermion systems is infrared divergent. One can remove these divergences by introducing counterterms. To do this without changing the model, one has to solve an inversion equation. We call this procedure Fermi surface renormalization (FSR). Whether or not FSR is possible depends on the regularity properties of the fermion self–energy. When the Fermi surface is nonspherical, this regularity problem is rather nontrivial. Using improved power counting at all orders in perturbation theory, we have shown sufficient differentiability to solve the FSR equation for a class of models with a non-nested, non-spherical Fermi surface. I will first motivate the problem and give a definition of FSR, and then describe the combination of geometric and graphical facts that lead to the improved power counting bounds. These bounds also apply to the four–point function. They imply that only ladder diagrams can give singular contributions to the four–point function.

---

# 1. Overview

In these two talks, I shall describe the foundations of the improved power counting bounds developed in [FST1], and show how to apply them to solve the problem of Fermi surface renormalization for systems with a nonspherical Fermi surface, such as electrons in a crystal. All results I will discuss are perturbative (this means analysis of all orders of perturbation theory), but some of them directly concern the singularities of the four–point function and thus have non–perturbative consequences. All results stated without an explicit reference are derived in collaboration with Joel Feldman (UBC, Vancouver) and Eugene Trubowitz (ETH Zürich) [FST1,2].

Before diving into the details, I want to discuss how these results fit into the general project pursued in our group. The motivation is twofold.

(1) Feldman, Knörrer, Lehmann and Trubowitz [FKLT] show that in a system with a dispersion relation $\varepsilon(\mathbf{p})$ that is not $T$–symmetric, i.e. where $\varepsilon(-\mathbf{p}) \ne \varepsilon(\mathbf{p})$ for all except a discrete set of momenta $\mathbf{p}$, the familiar Cooper instability is absent because of this asymmetry and the system behaves like a Fermi liquid. More precisely, they prove nonperturbatively that the *renormalized* perturbation expansion is analytic in a disk uniformly in the volume and the temperature. Thus this system is a Fermi liquid if the perturbation expansion can be renormalized and if renormalization does not change the model. In [FST1] and forthcoming papers [FST2], we prove that renormalization is possible and that it does not change the model. This is a necessary ingredient in the proof of the existence of that Fermi liquid. This existence proof does away with certain claims in the physics literature that no Fermi liquids can exist in two dimensions.

(2) In the filling region relevant for high–$T_c$ theory, the Fermi surface is nonspherical, in fact, quite near to nesting [K]. Although we will assume that no perfect nesting takes place in the sense that the Fermi surface is not allowed to have identically flat sides, one can learn from the way our bounds behave how things change as half–filling is approached. In other words, one can understand nesting effects in a fairly easy way using improved power counting. Precisely at half–filling, things work a bit differently, but we can also do renormalization.

Since the subject of many–body fermion theory goes back to the fifties, an obvious question is whether all this has not been done before. The answer is no, because the arguments provided in the literature came with some unproven, but plausible assumptions 'on the side'. This is not unusual in the physics literature, but some of these assumptions turned out to be very nontrivial. They concern the regularity (i.e. differentiability) of the fermion self–energy, which is assumed without proof. If you tend to believe that such things are no big deal and uninteresting for any real physics, let me remind you that a common feature of several proposals for non-Fermi liquid behaviour, e.g. [AIM], or the marginal fermi liquid scenario, is a fermion self–energy that is not a $C^1$ function of momentum and frequency. What we have done in [FST1,2] is to show that for systems in any dimension $d \geq 2$ with a sufficiently rapidly decaying interaction (e.g. $|\mathbf{x}|^{-d-3}$ will do), the fermion self–energy is indeed smooth enough such that non–Fermi liquid behaviour can only come from singularities in the four–point function. More precisely, this means that non–Fermi–liquid behaviour cannot be seen in any fixed order in perturbation theory. In contrast, the situation considered in [AIM] is that of fermions interacting with transverse gauge fields, which means long–range interactions, and there the first order self–energy correction already behaves as $p_o^{1/3}$ instead of $p_o$, so it changes the behaviour of the fermion propagator completely. The problem with a long-range interaction is much more singular than the one we have solved, but the case of a short–range interaction is already nontrivial from second order on because of the singularity of the fermion propagator.

## 2. Fermi Surface Renormalization

I will first describe a lattice system that falls into the class of models we consider, mainly to introduce the notation, and then give the more general hypotheses under which our theorems hold. Let $d \geq 2$ be the spatial dimension and $\Lambda \subset \mathbb{Z}^d$ be finite (any lattice $\Gamma$ of maximal rank can replace $\mathbb{Z}^d$ here). The case $d = 1$ has been treated by Benfatto, Gallavotti and coworkers [G]. Let $\mathcal{F}$ be the Fock space generated by the fermion operators satisfying the anticommutation relations

$$c_\alpha(\mathbf{x})c_{\alpha'}^+(\mathbf{x}') + c_{\alpha'}^+(\mathbf{x}')c_\alpha(\mathbf{x}) = \delta_{\alpha\alpha'}\delta_{\mathbf{xx}'} \tag{2.1}$$

The Hamiltonian $H(c, c^+) = H_o + \lambda V$ has a free part

$$H_o = -\sum_{\mathbf{x},\mathbf{y}\in\Lambda} t_{\mathbf{x}-\mathbf{y}} \sum_{\alpha\in\{\uparrow,\downarrow\}} c_\alpha^+(\mathbf{x})c_\alpha(\mathbf{y}) \tag{2.2}$$

which describes hopping from a site $\mathbf{y}$ to another site $\mathbf{x}$ with an amplitude $t_{\mathbf{x}-\mathbf{y}} = t_{\mathbf{y}-\mathbf{x}}$, and an interaction part with a small coupling constant $\lambda$. The interaction is a four–fermion interaction

$$V(c, c^+) = \sum_{\mathbf{x},\mathbf{y}\in\Lambda} n(\mathbf{x})v(\mathbf{x}-\mathbf{y})n(\mathbf{y}) \tag{2.3}$$

where

$$n(\mathbf{x}) = \sum_{\alpha\in\{\uparrow,\downarrow\}} c_\alpha^+(\mathbf{x})c_\alpha(\mathbf{x}) \tag{2.4}$$

is the number operator at $\mathbf{x}$. For instance, the simplest Hubbard model is given by $\lambda = \frac{U}{2}$ where $U$ is the usual Hubbard-$U$, and by $v(\mathbf{x}-\mathbf{y}) = \delta_{\mathbf{xy}}$, and the hopping term is $t_{\mathbf{x}-\mathbf{y}} = t$ if $|\mathbf{x} - \mathbf{y}| = 1$ and zero otherwise.

It will turn out soon that to do renormalization properly, we have to deal with a whole class of free Hamiltonians $H_o$. One general condition that we impose is that both the hopping term and the interaction have a spatial decay that makes their Fourier transform a $C^2$ function. This is fulfilled if the second moments are finite:

$$\begin{aligned}\sum_{\mathbf{y}\in\mathbb{Z}^d} |t_{\mathbf{x}-\mathbf{y}}||\mathbf{x}-\mathbf{y}|^2 &< \infty \\ \sum_{\mathbf{y}\in\mathbb{Z}^d} |v(\mathbf{x}-\mathbf{y})||\mathbf{x}-\mathbf{y}|^2 &< \infty\end{aligned} \tag{2.5}$$

At temperature $T$ and chemical potential $\mu$, the grand canonical partition function is given by

$$Z_\Lambda = \text{tr}\left(e^{-\beta(H-\mu N)}\right) \tag{2.6}$$

with $N = \sum_\mathbf{x} n(\mathbf{x})$, and $\beta = \frac{1}{k_B T}$. The trace is over Fock space. We want to determine whether observables, typically given by expectation values of polynomials in the $c$ and $c^+$,

$$\langle \mathcal{O} \rangle = \frac{1}{Z_\Lambda} \text{tr}\left(e^{-\beta(H-\mu N)} \mathcal{O}(c,c^+)\right) \tag{2.7}$$

have a finite thermodynamic limit and whether an expansion in $\lambda$ can be used to get their behaviour at small or zero temperature $T$. For instance, one would like to expand the two–point function

$$G_2(\mathbf{x},\mathbf{y}) = \langle c^+(\mathbf{x})c(\mathbf{y}) \rangle = \sum_{r=0}^{\infty} \lambda^r G_{2,r}(\mathbf{x},\mathbf{y}) \tag{2.8}$$

It is by now well-known that at $T=0$, $\lim_{\Lambda \to \infty} G_{2,r} = \infty$ for all $r \geq 3$ [FT1,FT2]. This infrared divergence makes renormalization necessary. It will turn out that renormalization is merely a method to do the expansion in a more clever way. For $T > 0$, the unrenormalized expansion is convergent, but the radius of convergence shrinks to zero as $T \to 0$ so that the result is not uniform in the temperature.

To study this problem, we use the standard path integral representation for the generating functional for the correlation functions. It is obtained by partitioning the interval $[0,\beta]$ into $L$ intervals of length $\beta/L$, and applying the Trotter product formula to the trace for $Z_\Lambda$ and $\langle \mathcal{O} \rangle$. In the limit $L \to \infty$, the trace becomes a functional integral over Grassmann fields $\psi(\tau,\mathbf{x})$ and $\bar\psi(\tau,\mathbf{x})$ with a new euclidean time $\tau$ varying from 0 to $\beta$. We also temporarily introduce an infrared cutoff $\varepsilon > 0$ to make the infinite volume model well-defined, by forbidding values $e(\mathbf{p}) < \varepsilon$. The generating functional for the amputated connected correlation functions,

$$e^{\mathcal{G}_\varepsilon(\chi,\bar\chi)} = \int d\mu_{C_\varepsilon}(\chi-\psi,\bar\chi-\bar\psi) e^{-\lambda V(\psi,\bar\psi)}, \tag{2.9}$$

is then a convolution of the Grassmann Gaussian measure $d\mu_C$, given by the propagator

$$C_\varepsilon(p) = \frac{1}{ip_0 - e(\mathbf{p})} 1(|e(\mathbf{p})| > \varepsilon), \tag{2.10}$$

with the exponential of $V$. Here the Matsubara frequency $p_0$ is the Fourier variable dual to $\tau$, and $p = (p_0,\mathbf{p})$. For $T > 0$, $p_0 \in \pi T(2\mathbb{Z}+1)$. For $T=0$, $p_0 \in \mathbb{R}$. The indicator function $1(E) = 1$ if $E$ is true and $1(E) = 0$ otherwise. Without the infrared cutoff $\varepsilon$, $\mathcal{G}(\chi,\bar\chi)$ would not be defined because of the infrared divergences mentioned above. The cutoff removes these divergences because it cuts off the singularity of the propagator $C$. When $\varepsilon \to 0$, the coefficients in the perturbation expansion of $\mathcal{G}_\varepsilon$ in powers of $\lambda$ diverge because it they are not yet renormalized. Renormalization will remove these divergences and we can then take the limit $\varepsilon \to 0$.

In the following, I set $T=0$ and $\Lambda = \mathbb{Z}^d$ to discuss the singularities that give rise to the infrared divergences. If $\lambda = 0$, the electrons are independent and one can calculate the correlation functions simply by doing a Fourier transform. For the lattice systems discussed above, Fourier space is given by $\mathcal{B} = \mathbb{R}^d/\Gamma^\#$ where $\Gamma^\#$ is the lattice dual to the position space lattice, e.g. $\Gamma^\# = 2\pi\mathbb{Z}^d$ for $\Gamma = \mathbb{Z}^d$. The Fourier transform of the hopping term gives the band structure (or dispersion relation)

$$e(\mathbf{p}) = -\sum_\mathbf{x} t_\mathbf{x} e^{-i\mathbf{p}\cdot\mathbf{x}} - \mu \tag{2.11}$$

which, for the Hubbard case, reduces to $e(\mathbf{p}) = -2t \sum_{i=1}^{d} \cos p_i - \mu$. At zero temperature, all states with $e(\mathbf{p}) < 0$ are filled. The boundary of the occupied region in $\mathbf{k}$–space is the Fermi surface

$$S = \{\mathbf{p} \in \mathcal{B} : e(\mathbf{p}) = 0\}. \tag{2.12}$$

The density $\rho = \frac{1}{|\Lambda|} N$ is given by the volume enclosed by $S$ ( in two dimensions by the area inside $S$). In the example of the Hubbard model, the function $e$ has its minimum at $\mathbf{p} = 0$, and it is strictly convex (and analytic) near this minimum. Consequently, the Fermi surface is strictly convex for $\mu$ slightly larger than $-2td$. This is the generic behaviour of systems in solid state models: the band function is strictly convex around a minimum, so a small (but macroscopic) occupation of electrons in that conduction band gives rise to a strictly convex, curved Fermi surface. As the filling increases (which happens if $\mu$ is increased), the shape of the Fermi surface changes and it even becomes diamond-shaped at $\mu = 0$ (half–filling) for the $H_o$ of the Hubbard model when $d = 2$.

Before discussing renormalization, I state our hypotheses more precisely. We assume that the Fourier transform $\hat{v}$ of the two–body potential $v$ is $\hat{v} \in C^2(\mathbb{R} \times \mathcal{B}, \mathbb{C})$, that

$$\hat{v}(-p_o, \mathbf{p}) = \overline{\hat{v}(p_o, \mathbf{p})}, \tag{2.13}$$

and that all derivatives of $\hat{v}$ up to second order are bounded functions on $\mathbb{R} \times \mathcal{B}$. Since $\lambda$ and $\hat{v}$ appear only in the combination $\lambda \hat{v}$, we may assume that $|\hat{v}|_2 \leq 1$, where $|f|_2 = \sum_{|\alpha| \leq 2} \|D^\alpha f\|_\infty$. Note that the interaction potential may depend on $p_o$ as well. To show convergence at large $p_o$ we need only that $\hat{v}$ approaches a finite limit as $p_o \to \pm \infty$ and that this limit is approached monotonically (see [FST2] for details). These assumptions about the behaviour at large $|p_o|$ are satisfied in all the models discussed above, they are in fact much weaker than the usual analyticity assumptions (if the interaction is instantaneous, as in the Hamiltonian used in the above motivation, its Fourier transform $\hat{v}$ is even independent of $p_o$). The physically relevant assumption is the regularity of $\hat{v}$ because it requires some decay in position space.

We assume that $e \in C^2(\mathcal{B}, \mathbb{R})$ and that for all $\mathbf{p} \in S$, $\nabla e(\mathbf{p}) \neq 0$. Moreover, we assume one of the following

(A)     $S$ has no identically flat sides (for a precise definition see [FST1])

(B)     $S$ is strictly convex with strictly positive curvature.

These assumptions exclude half–filling ($\mu = 0$) because there the Fermi surface has flat sides and because the gradient vanishes at the corners of the diamond. I will not discuss this case further here; I just remark that we can renormalize the expansion by different arguments [FST2]. Obviously, assumption $(B)$ implies assumption $(A)$.

The connection between the singularity of $C$ and the infrared divergences in the model is easy to see: the functional $\mathcal{G}_\varepsilon$ has an expansion in the fields,

$$\mathcal{G}_\varepsilon(\psi, \bar\psi) = \sum_{m \geq 0} \int \prod_{i=1}^{2m} dp_i \delta(p_1 + \ldots + p_m - p_{m+1} - \ldots - p_{2m}) G_{\varepsilon,m}(p_1, \ldots, p_{2m}) \prod_{i=1}^{m} \bar\psi(p_i) \psi(p_{m+i}) \tag{2.14}$$

with the kernels $G_{\varepsilon,m}$ given by a sum over values of Feynman diagrams. Every such contribution is a finite–dimensional integral. The integrand consists of various combinations and powers of $C$ given by the Feynman rules. However, powers of $C$ are in general not locally integrable: introducing variables $\rho$ transversal and $\omega$ tangential to $S$, the integral

$$\int\limits_{|ip_{\mathrm{o}}-e(\mathbf{p})|<\varepsilon} \frac{dp_{\mathrm{o}}d^d\mathbf{p}}{|ip_{\mathrm{o}}-e(\mathbf{p})|^\alpha} \underset{e(\mathbf{p})=\rho}{=} \int \frac{dp_{\mathrm{o}}d\rho}{|ip_{\mathrm{o}}-\rho|^\alpha} \int d\omega J(\rho,\omega)$$
$$= \int_0^\varepsilon \frac{rdr}{r^\alpha} f(r) \tag{2.15}$$

diverges for $\alpha \geq 2$. (Here $J$ is the Jacobian of the change of variables and $f$ its integral over $\omega$, so $f(0)$ is nonzero). By momentum conservation and the Feynman rules, graphs that contain a string of two–legged insertions produce arbitrarily high powers of $C$ and thus divergences (see [FST1] for a detailed explanation). The cutoff $\varepsilon$ removes these divergences, but the values of these graphs become large when $\varepsilon$ becomes small. To renormalize, we perform subtractions of the insertions at the singularity on $S$. The details of this subtraction can be found in [FST1]. We will discuss in detail below what these subtraction terms really are and give an alternative characterization which is less technical.

The physical intuition behind these divergences is very simple and has been well-known for a long time. One expects that the interaction produces a self–energy $\sigma(\lambda,p)$ of the fermion such that the propagator

$$G(p) = \int d\tau \sum_{\mathbf{x}} \langle \psi(0,0) \bar\psi(\tau,\mathbf{x}) \rangle e^{ip_{\mathrm{o}}\tau - i\mathbf{p}\cdot\mathbf{x}} \tag{2.16}$$

behaves essentially as

$$G(p) = (ip_{\mathrm{o}} - e(\mathbf{p}) - \sigma(\lambda,p))^{-1}. \tag{2.17}$$

If $\sigma$ is a reasonable function, then the integrability properties of $G$ will be the same as those of the free propagator $C$, but the singularity is at a different place, namely $e(\mathbf{p}) + \sigma(\lambda,0,\mathbf{p}) = 0$. Thus the Fermi surface moves when the interaction is turned on. An expansion

$$\int dp \frac{1}{ip_{\mathrm{o}} - e(\mathbf{p}) - \sigma(\lambda,p)} \to \sum_{r=0}^\infty \int dp \frac{\sigma^r}{(ip_{\mathrm{o}} - e(\mathbf{p}))^{r+1}} \tag{2.18}$$

just introduces artificial divergences. In other words, if one can put $\sigma$ into the denominator, the divergences should disappear.

There are two problems with putting $\sigma$ into the denominator. First, $\sigma$ is not known but has to be calculated itself. This is standard perturbation theory because the first order contribution $\sigma_1$ to $\sigma = \sum_{r \geq 1} \sigma_r \lambda^r$ is finite and one can therefore proceed recursively. The real problem is the assumption that $\sigma$ is a reasonable function. One has to show this to verify that the interacting propagator indeed has the same integrability properties as the free one. There is no assumption left free because $\sigma$ itself is determined by the free model and the interaction, so its regularity properties have to be proven. The same argument that shows finiteness of $\sigma_1$ also suggests that already $\frac{\partial}{\partial p}\sigma_2 = \infty$, even after renormalization. This is the behaviour suggested by naive power counting bounds. These bounds are not sharp, however, and in [FST1,2] we have improved power counting sufficiently to show that $\sigma \in C^{2-\epsilon}$ for all $\epsilon > 0$ if $d = 2$ and $\sigma \in C^2$ for $d \geq 3$.

We now proceed to do renormalization using counterterms instead of putting the $\sigma$ into the denominator. In our opinion, putting counterterms in the action makes the concepts clearer in this problem. Another reason to do that is that we can show more regularity of the counterterm function $K$ which essentially restricts $\sigma$ to the Fermi surface than of the self-energy $\sigma$ itself: we have shown that $K$ is $C^2$ for all $d \geq 2$. If one prefers to change the propagator, one should put $K$ in there instead of $\sigma$ to use our bounds.

One way to motivate putting counterterms is as follows: since turning on $\lambda$ makes the Fermi surface move, and since this movement causes all the trouble with the expansion, one can try to add a function $K(\lambda, \mathbf{p})$ to the bilinear part of the action such that the Fermi surface $S$ stays fixed. In other words, $K$ compensates all self–energy corrections that would move the Fermi surface under the interaction.

**Theorem 2.1** *Assume (A). There is a formal power series*

$$K^{(\varepsilon)}(\lambda, \mathbf{p}) = \sum_{r=1}^{\infty} \lambda^r K_r^{(\varepsilon)}(\mathbf{p}) \tag{2.19}$$

*such that the model defined as*

$$e^{\mathcal{G}^{ren}(\psi, \bar{\psi})} = \int d\mu_{C_\varepsilon}(\chi - \psi, \bar{\chi} - \bar{\psi}) e^{-\lambda V(\chi, \bar{\chi}) - \int dp \bar{\chi}(p) K^{(\varepsilon)}(\lambda, \mathbf{p}) \chi(p)} \tag{2.20}$$

*has Fermi surface fixed to $S = \{\mathbf{p} : e(\mathbf{p}) = 0\}$. Moreover, the kernels $G_{m,r}^{ren}$ of $\mathcal{G}^{ren}$ all have finite limits as $\varepsilon \to 0$, and $K^{(\varepsilon)}(\lambda, \mathbf{p})$ has a finite limit $K(\lambda, \mathbf{p})$. The Borel transform in $\lambda$ of $G$ has a positive radius of convergence uniformly in $\varepsilon$.*

Thus, fixing the Fermi surface indeed removes all infrared divergences. It is interesting to note that the counterterms are finite. This theorem [FST1] is a nontrivial extension of the statements proven in [FT1] because the counterterms are momentum dependent. The dependence of $K$ on $\mathbf{p}$ is really there in absence of rotational symmetry, and it leads to substantial technical complications. $K$ is also a functional of $e$, so $K = K(\lambda, e, \mathbf{p})$. Using (2.13), one can show that the self–energy $\sigma$ satisfies $\sigma(-p_o, \mathbf{p}) = \overline{\sigma(p_o, \mathbf{p})}$. This implies that $K(\lambda, \mathbf{p}) \in \mathbb{R}$ for all $\mathbf{p}$ because $K$ is constructed from the self–energy by evaluating at $p_o = 0$ and $\mathbf{p} \in S$. More technically speaking, the graphs contributing to $K$ are the two–legged one–particle irreducible graphs that also contribute to $\sigma$, but they are evaluated at $p_o = 0$ and $\mathbf{p} \in S$ (see [FST1], Section 2.3).

Although we have now removed the infrared divergences, we have done so at the price of changing the model. Because of the counterterm function $K$, the quadratic part of the action is now

$$\mathcal{A}_o = \int dp \bar{\psi}(p) (ip_o - e(\mathbf{p}) - K(\lambda, \mathbf{p})) \psi(p) \tag{2.21}$$

and corresponds to a free Hamiltonian with dispersion relation $e + K$, which is $\lambda$–dependent, instead of $e$. Thus, if $e$ is the free dispersion relation, Theorem 2.1 makes a statement not about the original model but about a changed model. To do the renormalized expansion for a prescribed free model with band structure $E$ and interaction $V$, one has to solve the equation

$$E(\mathbf{p}) = e(\mathbf{p}) + K(\lambda, e, \mathbf{p}) \tag{2.22}$$

for $e$. Equation (2.22) is the central equation of the problem. I first explain how we solve it and then show how to renormalize without changing the model by using the solution $e = \mathcal{R}(E, \lambda)$ of (2.22) ($e$ also depends on the two–body potential $\hat{v}$, but all bounds are uniform in $\hat{v}$ for the set of $\hat{v}$ specified above, so I suppress that dependence in the notation). Let

$$K^{(R)}(\lambda, \mathbf{p}) = \sum_{r=1}^{R} \lambda^r K_r(\mathbf{p}) \tag{2.23}$$

be the function $K$ up to order $R$ in perturbation theory.

Crudely speaking, the right hand side of (2.22) is the identity plus a small term, because $K^{(R)}$ is of order $\lambda$, so an iteration is the natural strategy to get a solution. However, because of the various dependences of $K^{(R)}$ on $e$ and $\mathbf{p}$ one has to be very careful what one means by small (the sum $f + \lambda g$ will have very different properties than $f$ if $g$ is a more singular function, no matter how small $\lambda$ is chosen). Since $e \in C^2$ is the basic condition for all our bounds, we need at least $K^{(R)} \in C^2$, because in every step of the iteration, $e$ gets replaced by $e + K$. Also, to use a fixed point theorem, one needs control over $\frac{\delta K^{(R)}}{\delta e}$. But since

$$\frac{\delta}{\delta e} \frac{1}{ip_\circ - e} \sim \frac{1}{(ip_\circ - e)^2}, \qquad (2.24)$$

taking such derivatives seems to lead to new divergences. Nonetheless we have the following theorem.

**Theorem 2.2** *Assume $(A)$. Then for all $R \in \mathbb{N}$, $K^{(R)}(\lambda, \cdot) \in C^1(\mathcal{B}, \mathbb{R})$, and $K^{(R)}$ is also $C^1$ in $e$. Denote the Fréchet derivative of $K$ with respect to $e$ by $\frac{\delta K^{(R)}}{\delta e} \in \mathcal{L}(C^1, C^0)$ and the sup norm on $C^0$ by $|\cdot|_\circ$. Then for all $h \in C^1(\mathcal{B}, \mathbb{R})$*

$$\left| \frac{\delta K^{(R)}}{\delta e}(h) \right|_\circ \leq \text{const } |\lambda| \, |h|_\circ. \qquad (2.25)$$

Because of (2.25), $\frac{\delta K^{(R)}}{\delta e}$ extends uniquely to a bounded linear operator from $C^0$ to $C^0$. The set of $e$ satisfying $(A)$ is open, so if $e_1$ and $e_2$ are close enough, $(A)$ holds for all $e$ on the line connecting $e_1$ and $e_2$. Then $e_1 + K^{(R)}(e_1) = e_2 + K^{(R)}(e_2)$ implies by Taylor expansion that $(1 + \mathbb{L})(e_2 - e_1) = 0$, where $\mathbb{L} = \int_0^1 dt \, \frac{\delta K^{(R)}}{\delta e}((1-t)e_1 + te_2)$. Since $R$ is fixed and $\frac{\delta K^{(R)}}{\delta e}$ is a bounded operator for all $t$, $1 + \mathbb{L}$ is invertible for $\lambda$ small enough. Thus, we have

**Theorem 2.3** *Assume $(A)$. For all $R \in \mathbb{N}$, there is $\lambda_R > 0$ such that for all $\lambda \in (-\lambda_R, \lambda_R)$, the map $e \mapsto e + K^{(R)}$ is locally injective.*

This implies uniqueness of the solution under the quite general conditions $(A)$. The existence proof requires the stronger assumptions $(B)$ because for that, we need to show that $K$ is even in $C^2$. It is a priori not clear that $K$ must have the same differentiability properties as $e$. One might be in the situation that one always loses some regularity, i.e. that $e \in C^k$ only implies $K \in C^{k-1}$, or that even $e \in C^\infty$ leads only to $K \in C^{k_\circ}$ for some fixed $k_\circ$. It took us some time and optimal bounds to prove that for $k = 2$, there is no loss of regularity.

**Theorem 2.4** *There is an open set $\mathcal{E} \subset C^2$ of dispersion relations fulfilling $(B)$ such that $K^{(R)} \in C^2$ for all $e \in \mathcal{E}$. There is an open subset $\mathcal{E}' \subset \mathcal{E}$ and for all $R \in \mathbb{N}$, there is $\lambda_R > 0$ and a map $\mathcal{R} : (-\lambda_R, \lambda_R) \times \mathcal{E}' \to \mathcal{E}$ such that for all $(\lambda, E) \in (-\lambda_R, \lambda_R) \times \mathcal{E}'$, $e = \mathcal{R}(\lambda, E)$ solves (2.22).*

Using Theorem 2.4, we can now do renormalization without changing the model, as follows. Let the model be given by a potential $V$ and by a dispersion relation $E$ for the independent electrons, with $E \in \mathcal{E}'$. Let $R \in \mathbb{N}$ and $\lambda \in (-\lambda_R, \lambda_R)$. Use Theorem 2.4 to determine $e = \mathcal{R}(E, \lambda)$, and set $\kappa(E, \lambda, \mathbf{p}) = K(\lambda, \mathcal{R}(E, \lambda), \mathbf{p})$. Then $E = e + \kappa = e + K(e)$. Denoting the propagator with $E$ by $C(E)$ and the one with $e$ by $C(e)$, we have, by standard shift formulas for Gaussian measures, the identity

$$\int d\mu_{C(E)}(\psi - \chi, \bar{\psi} - \bar{\chi}) e^{-\lambda V(\chi, \bar{\chi})} = \frac{Z_e}{Z_E} \int d\mu_{C(e)}(\psi - \chi, \bar{\psi} - \bar{\chi}) e^{-\lambda V(\chi, \bar{\chi})} e^{-(\bar{\psi} - \bar{\chi}, K(\psi - \chi))}$$
$$= \frac{Z_e}{Z_E} e^{-(\bar{\psi}, K\psi)} \int d\mu_{C(e)}(\psi - \chi, \bar{\psi} - \bar{\chi}) e^{-\lambda V(\chi, \bar{\chi}) - (\bar{\chi}, K\chi)} e^{(\bar{\psi}, K\chi) + (\bar{\chi}, K\psi)}.$$
(2.26)

This is an identical rewriting of the generating functional for the model given by $E$ and $V$ in terms of the quantities $e$ and $K$ that appear in the renormalized expansion, obtained by moving the $K$ from the propagator to the interaction. Since $E = e + K$, this leaves $e$ in the propagator. The change in normalization factor is irrelevant for any correlation function, and the extra source terms in the integrand just modify the external legs in a trivial way. This identity also holds if a cutoff $\varepsilon > 0$ is in place. In that case, all bounds are uniform in $\varepsilon$ and Theorem 2.1 implies that the kernels $G_{m,\varepsilon}^{(R)}$ converge as $\varepsilon \to 0$ (here the superscript $R$ indicates $G$ up to order $R$ in $\lambda$, similarly to the definition of $K^{(R)}$).

Physically, this procedure means the following. Applying the map $\mathcal{R}$, i.e. going from $E$ to $e$, shifts the Fermi surface from the free surface $S(E)$ to the interacting Fermi surface $S(e)$. Thus, in this step, the deformation of the surface caused by the interaction is taken into account. The renormalized expansion is then done at fixed interacting Fermi surface $S(e)$, and it can be used to calculate other self-energy effects, and the other correlation functions.

As mentioned in the statement of Theorem 2.1, the bounds that we prove for the kernels are not sufficient (and shouldn't be sufficient) to show convergence of the perturbation series in $\lambda$. This is the reason for the explicit restriction to a finite order in perturbation theory in the other theorems. The general bound we obtain is the standard de Calan – Rivasseau bound, e.g. for the two–point function

$$G_2(p) = \sum_{r=0}^{\infty} G_{2,r}(p) \lambda^r \tag{2.27}$$

it reads

$$|G_{2,r}(p)| \le r! \, Q^r \tag{2.28}$$

where $Q$ is some constant. If this bound is saturated, the perturbation series has convergence radius zero, and more precisely, it means that the $\lambda_R$ of the above theorems behaves as

$$\lambda_R \propto \tfrac{1}{R}. \tag{2.29}$$

The renormalization method yields statements about how and why these factorials can appear:

(1)  In order $r$, there are so many graphs that bounding the sum of graphs by the sum of their absolute values already gives an $r$ factorial. This bound is not sharp because in it, the Pauli principle is ignored. For fermions, one may expect sign cancellations, such as in determinant bounds, to be useful. Determinant bounds do not work uniformly in the cutoff in these models, however, and it is a hard problem to implement the Pauli principle to show that the number of graphs does not produce a factorial. This was done by Feldman, Magnen, Rivasseau, and Trubowitz [FMRT] for $d = 2$. A similar result is expected to hold for fermions in any dimension.

(2) singularities in values of individual four–legged diagrams can produce $r$ factorials as well. The best–known example of this are the BCS ladders which produce symmetry breaking [FT2].

For item (2), the improved power counting method provides the following theorem.

**Theorem 2.5** *Assume (A). The only graphs that can produce $r$ factorials are generalized ladder graphs.*

For details and the proof, see [FST1] and the next section. This theorem holds for any $d \geq 2$ and for the very general class of Fermi surfaces satisfying only the condition $(A)$ of non–flatness. No resummation, and hence no condition on the sign of the coupling is required.

The meaning of the term 'generalized ladder' is explained in detail in [FST1], and also below. The generalized ladders (called dressed bubble chains in [FST1]) are non–overlapping four–legged diagrams. By Lemma 2.26 of [FST1], any non–overlapping four–legged graph is a generalized ladder. It is constructed from the usual ladder graphs by replacing the bare vertices by effective vertices of a higher scale. The non–overlapping four-legged graphs emerge in a natural way in the renormalization flow because their scale behaviour is marginal, which produces the factorials. All contributions to the four–point function from overlapping graphs are bounded, or, in renormalization group language, irrelevant.

The importance of this is that convergence of perturbation theory can now be checked by looking at the ladders only: if they have singularities, then perturbation theory does not converge. If they don't, the expansion in $\lambda$ converges. The structure of the ladders is so simple that their properties essentially only depend on the fermion propagator, in particular, the Fermi surface. In absence of nesting (such as takes place at half–filling in the Hubbard model), the particle–hole ladders have no singularities. The particle–particle ladders always have a singularity at zero transfer momentum if the Fermi surface is symmetric, i.e. if $e(-\mathbf{p}) = e(\mathbf{p})$. The Fermi liquid of [FKLT] has a Fermi surface for which the particle–particle ladders are uniformly bounded as well because there is no symmetry of $e$ under $\mathbf{p} \to -\mathbf{p}$.

## 3. Improved Power Counting

In this section, I discuss the reasons behind the improved power counting bounds. I have written the present section so that it can be read as an easy introduction to the technical parts (Chapter 2 and 3) of [FST1]. I shall discuss two examples of graphs to bring out the main point. After that, it should be obvious to generalize it to all graphs, given the graph classification of [FST1], Section 2.4.

To do estimates we need some definitions from scale analysis. As in all modern treatments of renormalization, we decompose ('slice') the propagator around its singularity. There is a lot of freedom in doing this, but the decomposition is chosen such that the propagator has very simple behaviour on each slice. In the previous section, we introduced an infrared cutoff $\varepsilon$. Since all quantities scale like powers of $\varepsilon$ and $\log \varepsilon$, we take $\varepsilon$ of the form

$$\varepsilon = M^I, \quad \text{with } M > 1 \text{ fixed and } I \text{ a negative integer.} \tag{3.1}$$

Removing the cutoff thus means taking the limit $I \to -\infty$. Moreover, we will now trace back the behaviour when the energy scale varies by looking at the contributions from energy shells $M^{j-2} \leq |e(\mathbf{p})| \leq M^j$, for $I \leq j < 0$. This decomposition is natural because it is adapted to the singularity.

For definiteness, here are the details (if you are interested only in the main features of the decomposition, you can skip this paragraph). Let $r_o > 0$ be chosen such that in an $r_o$–neighbourhood of the Fermi surface the coordinates $\rho$ and $\omega$ of (2.15) can be used. Let $M \geq \max\{4^3, \frac{1}{r_o}\}$ (then $|e(\mathbf{p})| < M^{-1}$ implies $|\rho| < r_o$), and let $a \in C^\infty(\mathbb{R}_o^+, [0,1])$ such that $a(x) = 0$ for $0 \leq x \leq M^{-4}$, $a(x) = 1$ for $x \geq M^{-2}$, and $a'(x) > 0$ for all $x \in (M^{-4}, M^{-2})$. Set

$$f(x) = a(x) - a(\tfrac{x}{M^2}) = \begin{cases} 0 & \text{if } x \leq M^{-4} \\ a(x) & \text{if } M^{-4} \leq x \leq M^{-2} \\ 1 - a(\tfrac{x}{M^2}) & \text{if } M^{-2} \leq x \leq 1 \\ 0 & \text{if } x \geq 1, \end{cases} \tag{3.2}$$

so that, for all $x > 0$, $f(x) \geq 0$ and

$$1 - a(x) = \sum_{j=-\infty}^{-1} f(M^{-2j} x). \tag{3.3}$$

Calling $f_j(x) = f(M^{-2j} x)$,

$$\operatorname{supp} f_j = [M^{2j-4}, M^{2j}]. \tag{3.4}$$

The decomposition of $C$ is

$$\frac{e^{ip_o 0^+}}{ip_o - e(\mathbf{p})} = \frac{e^{ip_o 0^+} a(p_o^2 + e(\mathbf{p})^2)}{ip_o - e(\mathbf{p})} + e^{ip_o 0^+} \sum_{j<0} C_j(p). \tag{3.5}$$

where

$$C_j(p) = \frac{f(M^{-2j}|ip_o - e(\mathbf{p})|^2)}{ip_o - e(\mathbf{p})} = \frac{f_j(|ip_o - e(\mathbf{p})|^2)}{ip_o - e(\mathbf{p})}. \tag{3.6}$$

The first term in (3.5) is bounded and therefore cannot give rise to infrared singularities. A bit of care is required, however, to treat it in the ultraviolet (large $p_o$). This is not difficult and done in [FST2]. For the purposes of the present discussion, I just discard this term. Then all scale sums go from $j = I$ to $j = -1$, where $I < 0$ is the infrared cutoff, and all propagators are supported in a small neighbourhood of the Fermi surface. In particular, the $e^{ip_o 0^+}$ can in front of the sum in (3.5) can be omitted.

So much for the details which I included for definiteness. In the following we shall need only the following two facts: firstly, the cutoff propagator is a sum

$$C(p) = \sum_{j=I}^{-1} C_j(p), \tag{3.7}$$

and secondly, that the propagator on scale $j$, $C_j$, has simple properties: it is easy to prove (see [FST1], Lemmas 2.1 and 2.3) that

$$|D^\alpha C_j(p)| \leq W_s M^{-(s+1)j} \mathbf{1}(|ip_o - e(\mathbf{p})| \leq M^j) \tag{3.8}$$

where $D^\alpha$ is a derivative with respect to $p$ ($\alpha$ is a multiindex with $|\alpha| = s$, $0 \leq s \leq k$) of order $s$. The indicator functions take the value $\mathbf{1}(X) = 1$ if $X$ is true and $\mathbf{1}(X) = 0$ otherwise. In words: on 'slice' number $j$, the propagator is for all $(p_o, \mathbf{p})$ of almost constant absolute value, at most $M^{2-j}$, ($s = 0$ in (3.8); $W_o = M^2$), and each derivative produces another large factor $M^{-j}$. The constant $W_s$ depends on $|e|_s$ and on $g_o$ ($g_o$ is the lower bound on $|\nabla e|$ near the Fermi surface). Moreover, the support of $C_j$ is contained in the product of an interval of length $2M^j$ in $p_o$ and a thin shell $R_j$ of thickness const $M^j$ around the Fermi surface $S$. An example of such an $R_j$ is drawn for the two–dimensional case in Figure 1 $(a)$. By our choice of $M$, $M^{-1} < r_o$, so for all $j < 0$ this shell is contained in the region where the variables $\rho$ and $\omega$, introduced in (2.15), can be used.

### 3.1 Volume Improvement Bounds

The scale decomposition is a natural way to understand power counting because it allows us to weight the growth of the propagator in the vicinity of its singularity $S$ against the smallness of the volume of shells around the Fermi surface, where it becomes large. In our scale decomposition, momentum space is cut into shells around the Fermi surface, as sketched in Figure 1 (a). Because we have assumed that the gradient of $e$ does not vanish on the Fermi surface, the $\mathbf{p}$–volume of a shell, in which $M^{j-1} \leq |e(\mathbf{p})| \leq M^j$ $(j < 0)$, is bounded by a constant times $M^j$. Similarly, the support condition of the propagator $C_j$ restricts $p_o$ to the interval $[-M^j, M^j]$. One can deduce the integrability properties of the infrared part $(1-a)C$ of the propagator $C$ discussed above by weighting this volume against the growth of $|C|$ in a summation over shells:

$$\int_{\mathbb{R}\times\mathcal{B}} dp_o d^d\mathbf{p} \frac{1}{|ip_o - e(\mathbf{p})|^\alpha} (1 - a(p_o^2 + e(\mathbf{p})^2)) = \sum_{j<0} \int_{\mathbb{R}\times\mathcal{B}} dp_o d^d\mathbf{p} \frac{f_j(|ip_o - e(\mathbf{p})|^2)}{|ip_o - e(\mathbf{p})|^\alpha} \tag{3.9}$$

By (3.4), and because the size of $|C_j|$ is bounded by $M^{-j+2}$, this is bounded by

$$\sum_{j<0} M^{(-j+2)\alpha} \int dp_o\, 1(|p_o| \leq M^j) \int d^d\mathbf{p}\, 1_j(\mathbf{p}) \tag{3.10}$$

where $1_j(\mathbf{p}) = 1(|e(\mathbf{p})| \leq M^j)$. Since the volume of the $d$–dimensional shell

$$R_j = \{\mathbf{p} \in \mathcal{B} : |e(\mathbf{p})| \leq M^j\} \tag{3.11}$$

is bounded by const $M^j$, this is

$$\leq M^{2\alpha} \sum_{j<0} M^{-j\alpha}\, 2M^j\, \text{const } M^j$$
$$= \text{const} \sum_{j<0} M^{j(2-\alpha)} = \text{const} \frac{1}{1 - M^{-(2-\alpha)}} \tag{3.12}$$

if $\alpha < 2$. So, for instance the first order correction to the self–energy is a sum over scales $j < 0$ of

$$\sigma_{1,j}(q) = \int d^{d+1}p\, C_j(p)\hat{v}(q - p). \tag{3.13}$$

By the above bound with $\alpha = 1$, and since $\|v\|_\infty \leq 1$,

$$|\sigma_{1,j}(q)| \leq \text{const } M^j \tag{3.14}$$

so the sum converges. The scaling bound $M^j$ is the typical bound for self–energy contributions at a scale $j$. Note that this bound is not good enough to show convergence of a derivative of the self–energy because the derivative of the propagator $C_j$ contains an additional factor $M^{-j}$. If these bounds were saturated, the derivative of the self–energy would already be logarithmically divergent. For the first order, this is not so because the $C_j$ in (3.13) does not depend on the external momentum $q$, and because the interaction potential is $C^2$ by assumption. But in second or higher order, some of the propagators in the integral do depend on the external momentum and bounding derivatives becomes a problem, as I shall discuss below.

Up to this point, the scale decomposition introduced only a rewriting of (2.15) as a convergence statement for series instead of integrals. However, the geometry of the shells around the Fermi surface has important consequences for nontrivial graphs, which I discuss now. In the following, I impose the cutoff $I$ so that all scale sums run from $I$ to $-1$.

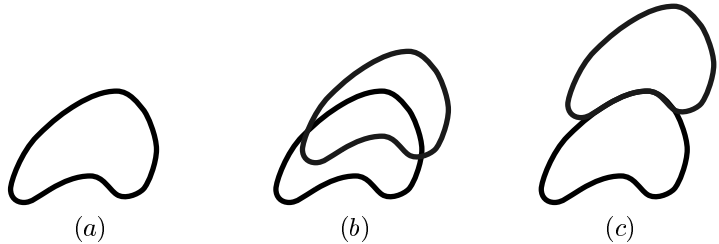

Figure 1: *Intersection of a shell around the Fermi surface with its translates*

I first start with a discussion of the contribution of the particle–particle bubble $B$ (shown in Figure 2) to the four–point function.

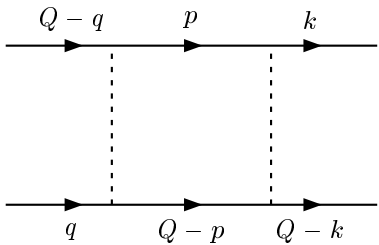

Figure 2: *The particle–particle bubble*

By momentum conservation, the sum of the ingoing momenta must equal that of the outgoing ones, so only three of the external momenta are independent. We choose them as indicated in Figure 2. Then

$$Val(B)(k,Q,q) = \int_{\mathbb{R}\times\mathcal{B}} d^{d+1}p\ C(p)\ \hat{v}(p-k)\ C(-p+Q)\ \hat{v}(q+p-Q). \tag{3.15}$$

Writing both propagators as scale sums one obtains

$$Val(B)(k,Q,q) = \lim_{I\to-\infty} \sum_{I\leq j,h<0} Val(B_{jh})(k,Q,q) \tag{3.16}$$

with
$$Val(B_{jh})(k,Q,q) = \int_{\mathbb{R}\times\mathcal{B}} d^{d+1}p\, C_j(p)\, \hat{v}(p-k)\, C_h(-p+Q)\, \hat{v}(q+p-Q) \qquad (3.17)$$

and therefore (since $|\hat{v}|_o \leq |\hat{v}|_2 \leq 1$)

$$|Val(B_{jh})| \leq M^{2-j}\, M^{2-h} \int dp_o\, 1(|p_o| \leq M^j)\, 1(|-p_o + Q_o| \leq M^h) \int d^d\mathbf{p}\, 1_j(\mathbf{p})1_h(-\mathbf{p}+\mathbf{Q}). \qquad (3.18)$$

Since $\mathcal{B} = -\mathcal{B}$, this bound is invariant under $\mathbf{Q} \to -\mathbf{Q}$ because the last integral is invariant under a change of integration variable $\mathbf{p} \to -\mathbf{p} + \mathbf{Q}$. Therefore, it suffices to discuss convergence properties of the sum over $I \leq j \leq h < 0$. We bound the $p_o$–integral and get

$$|Val(B_{jh})| \leq 2M^{2-j}\, M^{2-h} M^j V_1^{(-)}(\mathbf{Q}, j, h) \qquad (3.19)$$

where

$$V_1^{(\pm)}(\mathbf{Q}, j, h) = \int_{\mathcal{B}} d^d\mathbf{p}\, 1_j(\mathbf{p})1_h(\pm\mathbf{p}+\mathbf{Q}) = \text{vol}\, (R_j \cap (\pm R_h \mp \mathbf{Q})) \qquad (3.20)$$

is the volume of the intersection of shell $R_j$ with the translate of $\pm R_h$ by $\mathbf{Q}$ (see Figure 1 (b) and (c)). The only estimate uniform in $\mathbf{Q}$ is to bound this volume by that of a single shell,

$$V_1^{(\pm)}(\mathbf{Q}, j, h) \leq \text{const } M^j. \qquad (3.21)$$

I took $M^j$ here since $j \leq h$ implies $M^j \leq M^h$. With these bounds

$$\begin{aligned}|Val(B)| &\leq \text{const} \sum_{I \leq j \leq h < 0} M^{j-h} \leq \text{const} \sum_{j \geq I}\sum_{k \geq 0} M^{-k} \\ &\leq \text{const}\, \frac{1}{1-M^{-1}} \sum_{j \geq I} M^{0j} \leq \text{const}\, \frac{1}{1-M^{-1}} |I| \\ &= \text{const}\, |\log M^I|.\end{aligned} \qquad (3.22)$$

The marginal behaviour $M^{0j}$ is the standard power counting behaviour of four–legged graphs on scale $j$. This lack of decay causes a logarithmic divergence of the bound as the infrared cutoff $\varepsilon = M^I \to 0$.

In Figure 1 (b), one can see, however, that for those values of $\mathbf{Q}$ where the two surfaces intersect transversally, one has a much better bound because $\text{vol}(R_j \cap (\pm R_h \mp \mathbf{Q}))$ is only $\propto M^{j+h}$, instead of $M^j$ for an entire shell. But this volume gain is not uniform in $\mathbf{Q}$: it is certainly absent for $\mathbf{Q} = 0$. For the surface drawn in Figure 1, there are additional external momenta for which the intersection leads to no gain over the normal volume factor $M^j$, such as the one shown in Figure 1 (c). For strictly convex Fermi surfaces, one can prove

**Lemma 3.1** *If the Fermi surface has everywhere strictly positive curvature, then for $j \leq h$*

$$V_1^{(\pm)}(\mathbf{Q}, j, h) \leq \text{const } \begin{cases} \min\{M^j, \frac{1}{|\mathbf{Q}|}M^{j+h}\} & \text{if } |\mathbf{Q}| \leq \frac{k_F}{2} \\ M^j M^{h/2} & \text{otherwise.} \end{cases} \qquad (3.23)$$

*Here $k_F = \inf\{|\mathbf{p}| : \mathbf{p} \in S\}$.*

The extra factor $M^{h/2}$ comes from the curvature of the Fermi surface, which provides a volume gain for any $\mathbf{Q} \ne 0$. Lemma 3.1 implies the simpler bound

$$V_1^{(\pm)}(\mathbf{Q}, j, h) \le \text{ const } M^j \begin{cases} 1 & \text{if } |\mathbf{Q}| \le M^{h/2} \\ M^{h/2} & \text{otherwise.} \end{cases} \quad (3.24)$$

Inserting (3.24) into (3.19), and proceeding as in (3.22), one sees that for any $|\mathbf{Q}| \ne 0$, the sum converges. More precisely, we get

$$Val(B) \le \text{ const } |\log |\mathbf{Q}||. \quad (3.25)$$

The logarithm can be understood from (3.24): it is the scale $h_o = \max\{h < 0 : M^{h/2} \le |\mathbf{Q}|\}$. Below scale $h_o$, the convergent behaviour $M^{h/2}$ takes over. Above scale $h_o$, there is no decay in the sum, and that gives a logarithm in the same way as it appeared in (3.22), only that now $h_o$ replaces the cutoff scale $I$.

It is not hard to verify explicitly here that if $e(-\mathbf{p}) = e(\mathbf{p})$, this bound, obtained by looking only at volume effects, is sharp: $Val(B)(k, (0, \mathbf{Q}), q)$ diverges logarithmically at $\mathbf{Q} = 0$.

For the particle–hole bubble, where the lower propagator is in the other direction, the bound is not saturated because of a sign cancellation, and the function is finite at zero. This follows for $\mathbf{Q} = 0$ by Lemma 2.42 of [FST1], and for $\mathbf{Q} \ne 0$ by a Taylor expansion argument (see [FKLT]).

Physically, both these facts are important. The singularity of the particle–particle ladder causes Cooper pairing if the interaction is attractive. The absence of a singularity in the particle–hole ladder implies that there is no competition to Cooper pairing from the particle–hole channel. Note that Lemma 3.1 requires positive curvature. If the curvature vanishes, the particle–hole bubble can have singularities. This is, for instance, the case in the half–filled Hubbard model. As one approaches half–filling, the curvature tends to zero and a singularity builds up. It is believed to lead to antiferromagnetism exactly at half–filling. Note also that although the particle–hole bubble is bounded, its derivatives with respect to external momenta are not.

An analogous function, the polarization bubble, plays an important role in the second–order contribution to the self–energy. The polarization bubble is bounded, but not $C^1$ (actually, not even continuous) in the external momentum. Therefore it is a nontrivial question whether the value of the second–order skeleton self–energy correction shown in Figure 3 is differentiable or not.

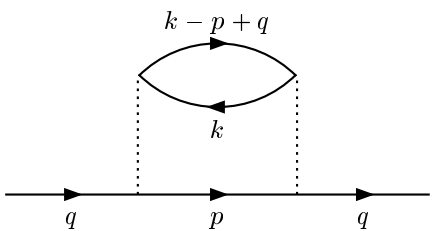

Figure 3: A second–order graph contributing to the self–energy

Its value is given by

$$Val(G)(q) = \int d^{d+1}p\, C(p) \hat{v}(q-p)^2 \int d^{d+1}k\, C(k) C(k - p + q) \quad (3.26)$$

The subintegral over $k$ is the polarization mentioned above. Note that this time a derivative with respect to $q$ will act on one of the propagators. Since the absolute square of the propagator is not integrable, we have to do the bounds carefully. Inserting scale sums as before, and bounding as before, we have for $i \leq j \leq h$

$$|Val(G)(q)| \leq \text{const} \sum_{i \leq j \leq h < 0} M^{-i-j-h} \quad M^{i+j} \quad V_2(\mathbf{q}, i, j, h) \quad (3.27)$$

$$\text{from } |C_j| \text{ etc.} \quad p_o\text{-integral} \quad \text{spatial volume}$$

where

$$V_2(\mathbf{q}, i, j, h) = \max_{u,v,w=\pm 1} vol\{\mathbf{p}, \mathbf{k} : |e(\mathbf{p})| \leq M^i, |e(\mathbf{k})| \leq M^j, |e(u\mathbf{p} + v\mathbf{k} + w\mathbf{q})| \leq M^h\} \quad (3.28)$$

is the integration volume for this two-loop graph. The restriction to $i \leq j \leq h$ is allowed because permutations of the scales only change signs in the linear combinations of the momenta, which we took into account by taking the maximum over signs $u, v, w$ in the definition of $V_2$. Again, the easiest bound for $V_2$ is obtained dropping the condition $|e(u\mathbf{p} + v\mathbf{k} + w\mathbf{q})| \leq M^h$, which gives

$$V_2(\mathbf{q}, i, j, h) \leq \text{const } M^{i+j} \quad (3.29)$$

From this, one can see that the sum for $Val(G)$ converges since it is majorized by

$$\text{const} \sum_{i \leq j \leq h < 0} M^{i+j-h} = \text{const} \sum_{i=I}^{-1} M^i \sum_{j=i}^{-1} \sum_{h=j}^{-1} M^{-(h-j)} \quad (3.30)$$

The sum over $h$ is a convergent geometric series bounded by $\sum_{k \geq 0} M^{-k}$, so (3.30) continues with

$$\leq \text{const} \sum_{i=I}^{-1} M^i \sum_{j=i}^{-1} \frac{M}{M-1} \leq \text{const} \sum_{i=I}^{-1} |i| M^i$$
$$\leq \text{const} \sum_{k \geq 0} k M^{-k} \leq \text{const.} \quad (3.31)$$

A derivative with respect to $q$ acts on the propagator $C_h$, and by (3.8), it produces another factor $M^{-h}$. Proceeding as above, one the series for $D_q Val(G)(q)$ is majorized by

$$\text{const} \sum_{i \leq j \leq h < 0} M^{i+j-2h} = \text{const} \sum_{i=I}^{-1} \sum_{j=i}^{-1} M^{-(j-i)} \sum_{h=j}^{-1} M^{-2(h-j)}. \quad (3.32)$$

The sums over $j$ and $h$ are convergent, but there is no decay left in the sum over $i$, and this suggests a divergence of the first derivative as $|I|$ when $I \to -\infty$.

However, the actual behaviour of $Val(G)$ is much better because of the following estimate for $V_2$. Written as an integral,

$$V_2(\mathbf{q}, i, j, h) = \max_{u,v,w=\pm 1} \int d\mathbf{p}\, 1_i(\mathbf{p})\, V_1^{(v)}(u\mathbf{p} + w\mathbf{q}, j, h). \quad (3.33)$$

Inserting the bound (3.24), we get two contributions,

$$V_2(\mathbf{q}, i, j, h) \leq \text{const } M^j \max_{u,w=\pm 1} \int d\mathbf{p}\, 1_i(\mathbf{p})$$
$$\left(M^{\frac{h}{2}} 1(|u\mathbf{p} + w\mathbf{q}| > M^{\frac{h}{2}}) + 1(|u\mathbf{p} + w\mathbf{q}| \leq M^{\frac{h}{2}})\right) \quad (3.34)$$

In the first term, we drop the **q**–dependent indicator function, so that the **p**–integral gives vol $R_i \leq$ const $M^i$. Thus

$$V_2(\mathbf{q},i,j,h) \leq \text{ const } M^j \left( M^{i+\frac{h}{2}} + \max_{s=\pm 1} \text{ vol } \left( R_i \cap \{\mathbf{p} : |\mathbf{p}+s\mathbf{q}| \leq M^{\frac{h}{2}}\} \right) \right) \tag{3.35}$$

The second term is the $d$–dimensional volume of the intersection of the shell $R_i$ of thickness $M^i$ with a ball of radius $M^{\frac{h}{2}}$ around $\pm\mathbf{q}$, sketched in Figure 4. This volume is bounded by the diameter of the ball times the thickness of the shell, which is at most const $M^{i+h/2}$, for all $d \geq 2$. For $d \geq 3$, the bound is even const $M^{i+h}$. We have thus shown the following volume improvement estimate.

**Lemma 3.2** *For all* **q**,
$$V_2(\mathbf{q},i,j,h) \leq \text{ const } M^{i+j+\frac{h}{2}}. \tag{3.36}$$

Note that, unlike the estimate for $V_1^{(\pm)}$, this estimate is uniform in the external momentum **q**.

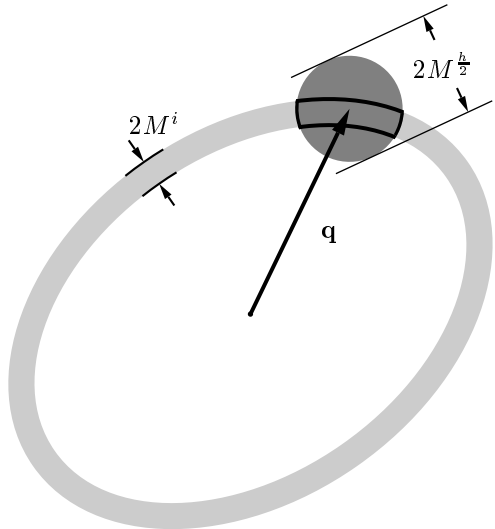

Figure 4: *The intersection of $R_i$ with the ball of radius $M^{\frac{h}{2}}$ around* **q**

Using this bound in the sum for $Val(G)$, we get a majorizing series

$$\text{const} \sum_{i \leq j \leq h < 0} M^{i+j-\frac{3}{2}h} = \text{ const} \sum_{i=I}^{-1} M^{\frac{i}{2}} \sum_{j=i}^{-1} M^{-(j-i)/2} \sum_{h=j}^{-1} M^{-3(h-j)/2} \tag{3.37}$$

which converges. So, $Val(G)$ is $C^1$, and in fact the remaining decay in the sum over $i$ can be used to show that its derivative is Hölder continuous in $q$ of order $\frac{1}{2}-\delta$ for any $\delta > 0$.

Above, I have only discussed boundedness of the scale sums uniformly in the cutoff $I$. Convergence of the functions as $I \to -\infty$ follows from the above bounds by an application of the dominated convergence theorem. For details, see Theorem 2.46 $(iv)$ of [FST1].

To summarize, we have shown for the second order graph of Figure 3 by some elementary volume estimates that for a strictly convex Fermi surface, the convergence of the scale sums is sped up such that its value is $C^1$ in the external momentum $\mathbf{q}$. We call this faster convergence improved power counting. It was shown in [FST1] that improved power counting holds for much more general Fermi surfaces than the strictly convex ones. They only have to satisfy the assumption $(A)$ mentioned above. The only difference is that instead of an improvement factor $M^{h/2}$ one gets in general only $M^{\epsilon h}$, with $\epsilon > 0$ depending on the surface. For example, the surface drawn in Figure 1 also has $\epsilon > 0$.

### 3.2 Non–overlapping Graphs

In this section, I generalize the improved power counting estimates to arbitrary graphs. This is possible because the entire argument leading to Lemma 3.2 used only very simple properties of the graph $G$, namely that $G$ has two loops that have a fermion line in common. We call such graphs *overlapping*. Whenever a graph is overlapping, the spatial volume integral appearing in the power counting bounds contains a subintegral bounded by $V_2(\mathbf{q}, i, j, h)$. For a large graph, the momentum $\mathbf{q}$ may be a very complicated combination of other loop momenta and the external momenta, but this is irrelevant because the volume improvement bound Lemma 3.2 is uniform in $\mathbf{q}$.

Therefore, any graph that contains two overlapping loops has an additional speedup of the convergence of scale sums. In general, we have if a two–legged graph is overlapping, its scale behaviour improves from $M^j$ to $M^{j(1+\epsilon)}$ with $0 < \epsilon \leq 1$.

if a four–legged graph is overlapping, its scale behaviour improves from $M^0$ to $M^{\epsilon j}$ with $0 < \epsilon \leq 1$.

The value of $\epsilon$ depends on the geometry of the Fermi surface. I have discussed only strictly convex surfaces here, but the volume gain from overlapping loops holds for much more general surfaces (see [FST1]), for example also for the one shown in Figure 1. The statements about the gain are true for overlapping graphs with an arbitrary number of external legs, but we need them only for two– and four–legged ones.

Obviously, now we only have to see which graphs are not overlapping, because all others have an extra gain that allows us to take an extra derivative in the two–legged case, and that gives us an extra decay that makes the value of $G$ converge, and free of singularities in the limit, in the four–legged case. For graphs with more than four legs, standard power counting is sufficient - we need only deal with the relevant ($E = 2$) and the marginal ($E = 4$) terms. Since we are only concerned with loops overlapping on fermion lines, we may replace interaction lines by vertices for the purposes of this graph classification. this replacement is indicated in Figure 5.
The classification of all non–overlapping graphs is done in Section 2.4 of [FST1]. That section of [FST1] is elementary and should be rather easy to read, therefore I will not discuss this any further here. The result is, as already anticipated in Theorem 2.5, the following.

**Theorem 3.3** *Let $G$ be a non–overlapping and one–particle irreducible graph, with vertices that have an*

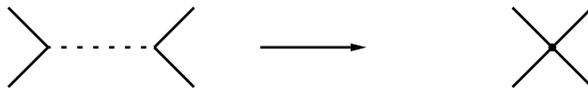

*Figure 5: Replacing interaction lines by vertices*

*even number of legs. If G is two–legged, G is a generalized Hartree–Fock graph. If G is four–legged, G is a dressed bubble chain.*

The exact definition of a generalized Hartree-Fock graph is given in [FST1] (there called GST graphs) and the dressed bubble chains are defined in [FST1] as well. Roughly speaking a generalized Hartree–Fock graph is built from a two–legged graph consisting of one vertex with self–contractions by (possibly recursively) replacing lines by other Hartree–fock graphs. Similarly, dressed bubble chains are built from simple bubble chains. These notions are probably easiest understood by looking at Figure 6. An ordinary Hartree–Fock graph would have only vertices with incidence number four. The essential difference between bubble chains and dressed bubble chains is that the dressed ones may have two–legged insertions of generalized Hartree–Fock graphs, and that the vertices may be effective vertices instead of bare ones. The restriction to one–particle irreducible (1PI) graphs in Theorem 3.3 is inessential since one–particle reducible graphs are strings of 1PI graphs in the two–legged case and, in the four–legged case, they are obtained by attaching a string of two–legged diagrams to each of the external legs of a 1PI four–legged graph (see [FST1], Remark 2.23).

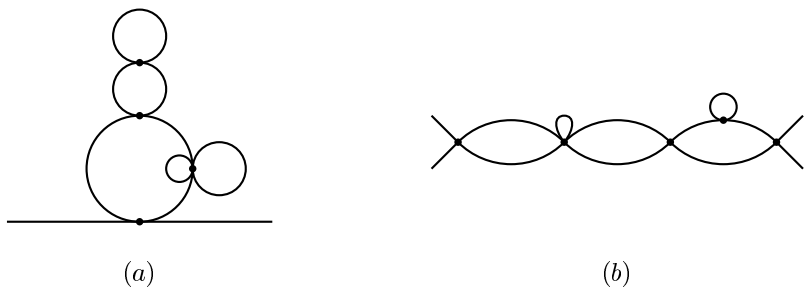

*Figure 6: (a) A generalized Hartree–Fock graph. (b) A dressed bubble chain.*

As is obvious from Figure 6, the structure of these graphs is simple enough for an explicit analysis. In particular, the differentiability problem for the self–energy is easy since the external momentum does not enter any line of a generalized Hartree–Fock graph, so that there is no problem of it acting on a line without an accompanying volume improvement factor arising at the same scale. Note that this holds only because all effective vertices have an even incidence number. For graphs that have vertices with odd incidence numbers, this statement would be false; see Remark 2.30 in [FST1].

For the four–legged graphs, the volume improvement bound gives a rigorous argument why one–loop summations take into account the leading behaviour: all other contributions are from overlapping graphs, hence nonsingular. In [FST1], this is formalized by introducing a number $n_f$ counting the number of times a non–overlapping four–legged subgraph can appear. The factorials in values of graphs are given by $n_\phi!$ where $\phi$ indicates the lowest scale in the graph (the root of the associated tree). See Theorem 2.47 of [FST1]. Some overlapping four–legged graphs are shown in Figure 7.

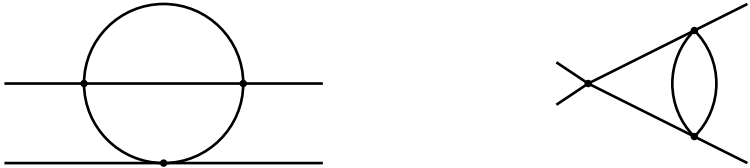

*Figure 7: Examples of four–legged overlapping graphs*

Two remarks are appropriate, however. First, the bubble chains of Theorem 3.3 are *not* the bare ladders of the theory: the vertices in the bubble chains can, and will, be effective vertices, i.e. their vertex functions are in general values of subdiagrams. The one–loop resummation justified by Theorem 3.3 is one where this scale dependence is taken into account, e.g. by defining suitable scale–dependent couplings. Similarly, the notion that a graph with scales associated to its lines is overlapping depends on the associated Gallavotti–Nicolò (GN) tree. All this is dealt with in [FST1]. Our graph classification does not require the vertices to be bare vertices. In particular, there is no requirement that each vertex have, e.g., at most four legs. The combination of graph classification with the scale structure leads to a natural decomposition of the GN trees, constructed in Section 2.5 of [FST1]. The number $n_f$ appearing in Theorem 2.46 of [FST1] is the number of bubble chains that appear in a graph associated to a particular GN tree, and it determines whether or not the value of a given $r^{th}$ order graph can have a value of order $r!$.

Second, the leading behaviour given by ladders determines whether or not singularities can build up because of momenta very near to the Fermi surface. It does not determine whether or not an effective attraction develops in a system with a repulsive bare interaction because the fields with momenta away from the Fermi surface may also produce such an attraction. In particular, when the curvature is not uniformly bounded below as a function of the filling, the ladders may start to dominate only at rather low scales and the effective interaction just above these scales has to be calculated using different justifications for the approximations. This is very important for a correct analysis of the effect of nesting of the surface on induced attractive interactions.

### 3.3 Further remarks

I end with a few remarks that relate the results stated in Section 2 to the bounds motivated in Section 3. I start with the discussion of the regularity of the self–energy. The above bounds were only sufficient to prove that the self–energy and the counterterm function $K$ are $C^1$ in the external momentum. How does one get from that to the statement that $K$ is $C^2$? It turns out that getting $C^2$ is still a much trickier problem which cannot be done by the above volume estimates alone. The main points leading to its solution are:

1) For the set of Fermi surfaces satisfying $(B)$, the optimal volume improvement exponent (shown to be $\geq \frac{1}{2}$ above) is actually 'almost equal to one', more precisely, the gain in Lemma 3.2 is not just $M^{h/2}$ but $|h|M^h$ in $d=2$, and $M^h$ in $d \geq 3$. Inserted into the bounds, this shows that the second derivative of $\Sigma$ and $K$ is at most logarithmically divergent, so one may hope for convergence by doing still more careful bounds.

2) In second order, this is possible by a very detailed analysis of the singularities that goes well beyond volume bounds, and one can even show that $K$ is $C^k$ if $e$ is $C^k$, for any $k \geq 2$. For $d \geq 3$, we can show that the self–energy $\sigma$ is $C^2$ if $e$ is $C^2$. In two dimensions, our bounds for the second derivative of $\sigma$ are logarithmically divergent, i.e. $\sigma \sim k_o^2 \log k_o$ for $k_o \to 0$. Explicit calculations [F] show that there is indeed such a logarithm, so our bounds do not overestimate the actual behaviour. Here one sees that the function $K$, which is $C^2$ also for $d=2$, is more regular than $\sigma$.

3) The analysis of the second–order $K$ is quite tricky, and it depends on too many details specific to the second order situation to be done for general graphs. However, by a new graph classification involving double overlaps, we can show that volume estimates are sufficient to prove finiteness of the second derivative for all graphs except the second order and two related graphs (which can also be treated explicitly) [FST2]. This also provides a classification of all graphs that can contribute to the $k_o^2 \log k_o$–behaviour of $\sigma$. The graphs that contribute can be obtained by a generalized RPA resummation.

The combination of these items into a common strategy is done in [FST2].

The reason why the asymmetric system is a Fermi liquid is now rather easily discussed: The asymmetry of $S$ implies that the function $V_1^{(-)}$ obeys a bound $M^{j+h/2}$ uniformly in $\mathbf{Q}$ [FKLT]. In other words, at $\mathbf{Q} = 0$ there is no singularity because $S$ and $-S$ are transversal to one another. There is also no singularity at nonzero $\mathbf{Q}$ because there there is always a curvature effect that gives a little extra volume gain, as for nonzero momenta in Lemma 3.1. Therefore there are no singularities in the ladder diagrams. By Theorem 3.3, no other contribution to the four–point function can be singular, so the four–point function is bounded. This is converted into a full, non–perturbative, proof of convergence in [FKLT].

# 4. Summary

In summary, improved power counting is a method to determine the regularity of Green functions, and thus of the self–energy, the vertex function etc., to all orders of perturbation theory. It is based on elementary volume estimates that follow directly from the geometry of the Fermi surface. These bounds do not require any explicit calculations, such as doing frequency sums or angular integrals exactly, but they are sharp, that is, they give the correct behaviour of the self–energy and the particle–particle Green function. One can see this by comparing our bounds (which apply to arbitrary graphs) to explicit calculations in low orders. The only case known to me where the bound overestimates the actual behaviour is that of the particle–hole bubble. In that integral there is an extra sign cancellation, which is well–understood (also in presence of a cutoff, see Lemma 2.42 of [FST1]). The notion of overlapping graphs makes precise in which situations certain one–loop resummations are justified. The condition that overlaps, or even multiple overlaps, occur in large graphs provides a rigorous meaning to the statement that 'more loop integrations smooth out the singularities'. The graph classification of [FST1,2] determines for which graphs this smoothing really occurs: there is no smoothing for non–overlapping graphs.

The geometric conditions we make on the Fermi surface are not only sufficient, but also necessary for these regularity properties. For surfaces with flat sides, the volume improvement effect, and thus the power counting gain discussed above, is absent, and indeed, the singularities suggested by naive power counting are really there. For instance, the self–energy fails to be $C^1$.

Finally, let me mention again that the strict convexity assumption $(B)$ was needed only to prove the existence of $\mathcal{R}$. In particular, Theorem 3.3, as well as all other results of [FST1] apply to any situation where the Fermi surface is not identically flat. We hope to prove the existence of the map $\mathcal{R}$ also for more general Fermi surfaces, e.g. where the curvature may change sign.

**Acknowledgements:** I would like to thank Joel Feldman for carefully reading this text and for various helpful suggestions. I also thank Walter Metzner for pointing out reference [F] to me, Volker Bach and the Erwin–Schrödinger Institute for the invitation to this very enjoyable workshop, and the ESI for its support during the workshop.